# Virtual and Augmented Realities as Symbolic Assemblies

Charles Bodon
bodonbruzel@gmail.com

**Abstract**

Against all attempts that consider virtuality as a substance (a parallel or alternative reality) or as a modality (like potentiality or possibility), we want to defend the pragmatic point of view that it is rather a dynamic cognitive and sensitive interaction with reality. More precisely, we show that the "*virtus*" is an operating capacity that produces simulations of real and fictional contexts to experiment with their effects. Based on Peirce's semiotics, we define virtual reality (VR) and augmented reality (AR) as mixed realities made of "symbolic assemblies", that is to say, structures of signs assembled by processes of computation and meaning (*semiosis*). We show that VR can be defined as a synesthetic experiment that does not reshape reality itself, but rather the senses and understanding we already have about it. In conclusion, we criticize David Chalmer's extended mind theory by distinguishing between knowledge and information, and we try to redefine AR as a hermeneutic device that extends not the mind itself, but the activity of thought by adding symbols to read in the world.

# Introduction

First of all, I warmly thank the University of Turin and Pr. Maurizio Ferraris for their invitation, it is a great pleasure for me to be here and present my research.

Nowadays, technology plays a central role in our lives. Every day we use smartphones to communicate and share information, while AI continues to improve by enhancing automation and mimicking our cognitive functions. More, recently, these technologies have begun to reshape our relationship with reality.

Indeed, Virtual Reality (VR) devices, such as VR headsets, can immerse the user in a digital environment, either real (such as a 3D model of Notre-Dame de Paris) or fictional (such as the planet Azeroth in the videogame *World of Warcraft*), while Augmented Reality (AR) overlays digital elements onto the real world through devices like smartphones, glasses, or QR codes.

So, one of the main differences between VR and AR can be determined in terms of *immersiveness*: where VR offers full immersion, AR is only partial. But both raise the same question: *does this immersiveness enrich or degrade reality?*

At first sight, these devices seem to degrade the reality for their users. Indeed, when they are combined with other technologies (such as generative AI or Deepfake) they can produce mental illness (such as derealization, social isolation) or increase the propagation of fake news by creating virtual items almost impossible to distinguish from real ones.

But, on the other hand, VR enables realistic simulations and the creation of fictional worlds, while AR enhances social interactions by connecting people through digital layers and providing information on objects. In this sense, these technologies have a scientific and artistic interest, but also a social utility that is improving the reality of their users.

So, we see that answering the question is not easy and how the ice can become thin between utopia and dystopia. On one side, VR and AR are concealing reality from us by provoking an ontological confusion between virtual and real objects. But they are diminishing it precisely by creating *new kinds of realities*.
So, one could say that VR and AR are also re-enchanting reality. But, of course, the way they did it is through simulations and fictions. So, if these technologies are improving reality, it is more exactly by creating *new kinds of artificial paradises*, which is a disappointing answer as well.

However, I think that we can get out of this dilemma if we focus on what these devices *do* more than their presupposed nature.

What I want to defend in this communication is that VR and AR are not in competition with reality nor re-enchanting it, even if they are producing fictions. More precisely, what I want to show is that they are only *manners* to interact and interpret reality. That is to say, manners that are concerning our *sense* of reality but not reality *itself*.

To do that, I will propose an approach that aims to define VR and AR as *mixed realities* made of what I call "symbolic assemblies". That is to say, hybrid structures made of signs and assembled by cognitive processes and sensations. To be immediately clear, I am not going to say that virtual entities are just existing "symbolically" in a metaphorical way that is hypnotizing us. On the contrary, what I want to show is that these entities are real and concrete, and that these immersive technologies depend on how we use them, more than they use us.

To do so, I will present my talk in three main points.

First, I will define what the word "virtual" means according to Charles Sanders Peirce's semiotics and try to show that, originally, virtuality is not an alternative reality, but rather a kind of cognitive interaction with it.
Then, I will show how this cognitive definition of virtuality is applicable to the context of virtual reality in informatics, and how we can redefine virtual reality as a *synesthetic* experiment.
And, finally, following this approach, I will discuss the case of AR and criticize David Chalmer's idea, according to which AR extends our mind in a *symbiotic* way.

# 1. Conceptual Definition of "Virtual"

## 1.1. History of the Concept and its Ambiguity

I start now with my first point and I want to define what "virtual" is from a conceptual and practical point of view.

Historically, the word "virtual" comes from "*virtualis*" or "*virtus*" which is a Latin translation from Thomas Aquinas of the Aristotelian concept of "*dunamis*" which can mean "potentiality", "capacity", or "possibility".

And, indeed, in the ordinary language, "virtual" is often a synonym of something that exists but in a potential way. The traditional example is the one of the tree virtually contained in the seed. Here, one says that the tree exists in the seed because it has the potentiality to grow from it, but it is not yet actualized, so it is virtual.

But here, we can also see the origin of the ambiguity of the word "virtual", because a virtual object seems to both exist and not exist simultaneously. For example, the virtual tree exists as potential in the seed, but doesn't actually exist because it hasn't grown up yet.

So, the word "virtual" is ambiguous because it implies the idea of something existing on the edge of two modalities of reality that are contradictory together: potentiality and actuality. And it is this ambiguity of virtuality that often led common sense or science fiction to substantialize virtuality as a parallel or alternative reality.

However, the problem with this metaphysical conception of virtuality is that it is a reification considering it as a "third realm" where the existence of virtual things would be hesitating between being real and unreal, potential and actual, which, of course, shocks the principle of contradiction.

But I believe that we can extricate ourselves from the *matrix* of such reification of virtuality if we cease to consider it as a substance or a modality and focus more on what kind of action it is.

In fact, virtuality is something that we do all the time when, in order to face a situation or to experiment with an object (real or fictional), we create what I call "symbolic assemblies", that is to say, *simulation of real or fictional objects and situations in order to provide us real effects*.

## 1.2. Peirce's Semiotics Conception of Virtuality

To defend and explain this position, and clarify how this concept of "symbolic assembly" is connected to virtuality, I will base my ontology on the semiotics approach that Charles Sanders Peirce give about the activity of thinking and his definition of "virtual" which I must briefly expose now.

For Peirce, thinking isn't an internal process confined to the brain, but rather a logical activity that operates through external signs to create abstract models. This process, called 'semiosis', forms meaning through a triadic relation between a sign (or 'representamen'), an object, and an 'interpretant', which is the cognitive effect produced by connecting the sign to the object.

For example, when I drive, if I see a red light, I connect this visual sign to the traffic light and the cognitive effect of this connection *prompts* me to stop my vehicle. So, the sign of the red-traffic light has the potential to stop me because of the cognitive effect that my interpretation of it produces.

And here is where we can connect this example to the definition Peirce gives about "virtual". As he wrote it in the Collected Papers CP. 6.372, §11, p. 4261:

> "A virtual X (where X is a common noun) is something, not an X, which has the efficiency (*virtus*) of an X."

In other words, a virtual X can *have* the effect of a real X, but without *being* that X. That is to say, here, Peirce makes the difference between "*having* the effect of something" and "*being* the thing that has an effect".

For example, my interpretation of the red light is a virtual "stop", because the red light does not physically stop me like an obstacle or a force blocking or holding my car would. However, this "virtual" stop *has* the same effect on me as a real one, because when I see it, I do really stop. In this sense, my interpretation of the red light has the *virtus*, that is to say the *efficiency* of a real stop without *being* an actual one.

So, for Peirce, the thinking process is fundamentally virtual, because it consists of producing abstract or formal configurations of a contextual situation by connecting signs together, and whose connection gives me real cognitive effect.

Anything—a person, object, event, or sensation—can be a sign, but its meaning will depend on the context in which it is perceived and used. And this point is important, because it means that, for Peirce, these semiotic relations are not fixed. Indeed, their meaning can evolve and give me new effects, according to the situations and combinations I do with other signs.

For example, I can reuse the interpretation I have of a red signal in different contexts as a rule that means that seeing a red signal implies stopping or canceling any kind of action (for example, turning off my computer by pressing the red power button). As well, the interpretation I have of a fictional character (let's say Ulysses from the Odyssey) can evolve as I read (for example, the abstract model I have of Ulysses is not the same when he fights Polyphemus, meets the gods, or returns to Ithaca).

In other words, for Peirce, there is a generativity of meaning, because a sign virtually contains multiple possible meaning and its interpretation actualize one of them. In this sense, the semiosis process is continuously open and virtually infinite.

## 1.3. The Virtuality of the Mind and Definition of Symbolic Assemblies

Now, we can see more clearly why virtuality is not as a substance in contrast with reality, but rather a cognitive and sensitive interaction with it through signs. When I perceive a red light or when I read the Odyssey, I produce virtual models of a contextual situation or object (even fictional) by connecting signs together. These models are not real objects or situations, but their interpretation has a cognitive effect on me that help me to interact with reality in any kind of context.

Thus, what I call a "symbolic assembly" is following this semiotic conception of Peirce:

> "A symbolic assembly is a combination of signs that produce a simulation of an X (real or fictional) which is not an actual X, but that has its efficiency, that is to say, whose interpretation can produce a real effect."

And I use here the term "assembly" on purpose, instead of "construction" or "agency" for example, because an assembly implies the idea of an interaction with an incarnate structure that is not fixed and whose shape can evolve by manipulation and association of its elements.

In this context, the *virtus* of an object can also be defined as *its capacity to produce cognitive or perceptual effects comparable to those that a real or fictional object would produce*.

With this definition, we can also answer the question of why virtuality seems often at the edge of possibility and actuality. The reason is that virtuality is a specific *sense* or *manner* of the mind to experiment with the effect of an object or situation (the red-traffic light or my mental picture of Ulysses) *as if* we were actually facing it.

In other words, virtual objects are a kind of counterfactual and if they are fictional, they are not fake as they have a functional reality by the effects they produce. In this sense, we can say that virtuality is not the same thing as reality, but rather an extension of it that we open to explore the space of its potentialities[1].

## 2. Virtual Reality as Synaesthesis

Now I have settled my definition of virtuality, what I want to show, for my second point, is that virtual processes in informatics are fundamentally an operationalized version of this cognitive activity of manipulating signs and producing simulations.

I want as proof that virtual items, worlds, characters, objects, etc., but also text editors, emojis, or videos, are ultimately produced by programs that read, write, and compute binary digits. Also, programs such as artificial intelligence are, in Turing's own words, "symbolic systems".

---

[1] CP 8.248 Cross-Ref: †† 248., "I do not say that we are ignorant of our states of mind. What I say is that the mind is virtual, not in a series of moments, not capable of existing except in a space of time -- nothing so far as it is at any one moment. †3", p. 4887

That is to say, mechanical procedures that consist of writing, reading, and operating signs together.

So, according to this definition, VR environments and items are also simulations, that is to say *complex symbolic assemblies that simulate virtual objects that give the effect of real or fictional objects without being ones*.

But what I want to emphasize now is that in order to produce the *virtus* of their simulation, VR programs must align two symbolic processes: i) first, the program's computations that generate virtual items and environments, ii) and, second, the user's semiosis process that generate meaning for these virtual programs.

In other words, what I want to show now is that VR programs are not altering the sense I have of reality, as we often hear it. On the contrary, for me, to be efficient VR programs *depends on* the understanding and feeling I already have about reality.

I want as my first argument the fact VR environments and items depend on my interactions with them in order to *exist*.

For example, in a video game, if my avatar leaves a castle and walks away, then the processor of my computer will simply stop computing its program or change its properties. The castle will progressively disappear in the long distance and literally cease to exist as I walk away, or reappear as soon as I approach to its zone, and its structure will change if I attack it.

In this sense, VR items and environments are not fixed symbolic assemblies, because the programs need to be in a perpetual interpretation of the multiple possible effects their users could generate according to their actions. Therefore, here, the definition I gave earlier about virtuality as *a generative space of potentialities extracted from contextual situations*, seems to be also applicable to an informatics context.

For my second argument, I want to remind us that VR programs are also dependent on my interpretation of them in order to have *meaning*.

Indeed, to provoke my immersivity, the simulation needs to be able to stimulate my senses by reproducing accurately *how they would have perceived real effects* such as the depth of the sound, spatial dimensions, and how I understand causal logic.

For example, if I pass through a virtual wall meant to block me, or throw a rock against a window that doesn't break it, my immersion breaks due to the contradiction between my actions and their expected effects.

And this dependency on my interpretation also works if I'm in a fictional virtual world. For example, suppose I am in a virtual Star Wars universe. In that case, the program has to be coded in order to be able to reproduce phenomena that are meaningful in this universe, such as the ability to move objects by the Force, that is to say without touching it.

In other words, *a virtual fictional world's meaning also depends on its fictional context's interpretative norms*.

Thus, VR programs are symbolic assemblies that I interact with and that give me *real sensations*, but that I also interpret in order to give them a *contextual signification*.
They are both interactive and hermeneutics systems that aim to make correspondences between the senses of its user and meaningful structure of signs. In other words, they provide a *synesthetic experiment* to their user in which he is actively participating.

## 3. The Augmented Mind Hypothesis: Augmented Reality as Hermeneutics

Now I have established that VR is not an alternative reality but rather a technic that assembles miscellaneous signs that compose not reality *itself*, but rather my *sense* of reality, I will consider, for my final point, how this symbolic activity unfolds in the less immersive context of AR.

As David Chalmers wrote in his book *Reality+*, Chapter 16, p. 294: "*AR simultaneously augments the world and augments the mind*".

For him, AR devices such as Google glasses are not just exteriorizations of our cognitive activity, an "*exocortex*" as he calls it, but literally become part of our mind as they progressively take over function of the brain such as memory, spatial navigation, language processing, image recognition, etc.

By playing on words, we could say that for Chalmers, in the opposite of VR where the user's mind is immersed in a technological device, with AR it is rather the device that is immersed in its user's mind whose everywhere outside. So, in fact, it is a kind of internalism but "externalized".

Therefore, for Chalmers, the process of what he called the "extended mind", the hypothesis according to which cognitive processes are external and that technical objects are extending them, becomes a reality with AR. For Chalmers, we are slowly, but surely, turning ourselves into cyborgs, that is to say, organisms in a symbiotic relationship with technology.

I don't completely disagree with Chalmers about the idea that we are becoming increasingly dependent on technology. However, I take issue with the terms 'extension' and 'augmented' when applied to the mind.

Indeed, if we take this expression seriously, we can see that it is grammatically incorrect because it treats the mind as a physical space that can be expanded or as a storage processor that could gain more power, which is a misleading metaphor.

For me, the problem in Chalmers' conception is that it considers that AR devices provide the *same* or *more* knowledge about the world as natural cognitive functions do. His argument is the following, p. 302:

1. Since cognitive functions are genuine knowledge,
2. And that external functions of AR devices can play the same role as natural cognitive functions,
3. Therefore, external functions of AR are genuine knowledge.

I think this syllogism is false because what AR devices produce is not *knowledge*, but *information*, and this is not the same thing.

Indeed, the difference we can draw between both is that knowledge is something that you have or possess because you learned it, while information is something that it is temporarily given to you about something and you don't own.

In other words: *information is what is given to you but that you don't own, while knowledge is what you own but that is not given.*

And I want as an example the same story that Chalmers used in his book. He shares the story of Charles Stross' novel called "*Accelerando*" where the main character, Manfred Macx, wears AR glasses that store his memory and gather information for him about the world.

The glasses help him to do every task of his day, such as finding his way to navigate in the city or describing objects by adding signs and information on them. But in the story, they are eventually stolen and Manfred Macx is almost helpless because he loses his memory and all the information about the world he had stored since he used these glasses.

Ironically, this story is a contradiction to Chalmers' argument, because it exactly shows that information is not *in* the mind nor that it extends it because when you remove or lose the AR device you immediately lost all of your data.

However, this is not the case with knowledge. For example, when I know how to go by car to the opera, I don't lose this knowledge. Even if I struggle to remember the road, it is still something that I can remember because I have learned it through practice. With an AR device (or a GPS) I don't practice anything by myself, but I follow instructions.

Therefore, we are not in a symbiotic relationship with AR devices, because the information they provide is not (yet?) *in* our mind or body nor it augments them like the storage of a computer's memory.

However, what AR certainly increases is our interpretative activity. Indeed, AR produces more signs for us to read in the world. Therefore, by doing so, this symbolic production does not *augment* the mind itself, but rather *extend* the *domain of thought*.

Thus, I think it is more realistic to join here the documental model of the mind that Maurizio Ferraris developed in *Documentality*, 5.3, p. 284. Indeed, for Maurizio Ferraris, what we call "spirit" is a social product that is emerging from an activity of registration, and artifacts, such as AR, serve as producing registrations and signs that extend the *social sphere*.

## Conclusion

Because it is time for me to conclude, I would like to summarize the main ideas I presented.

I have shown that virtuality, using Peirce's semiotics, is not an alternate reality, a substance, or even a modality, but the cognitive process of semiosis consisting of producing abstract models, what I called "symbolic assemblies", of contexts by connecting their signs.

I believe this position to be realistic, because this is actually, as I showed in my second point, what we also do with virtual reality in informatics by producing digital simulations through symbolic processes. I also tried to overturn the common idea that VR pushes its user toward derealization and, instead, showed that it was a synesthetic experiment that stimulates and depends on our senses and understanding of reality.

Finally, it leads me to discuss the case of AR and what I call "Chalmer's symbiotic conception" of the relationship between technology and the mind. By making a distinction between knowledge and information, I tried to show that we need to be precise when we say that "AR extend the mind and the world". What AR devices really do is that, by extending the domain of reading, they augment not the mind itself, but rather extend its *presence* in the world.

In each case, I tried to show that virtuality doesn't transfigure reality, but rather is a mixed reality involving senses, intellect, and sociality. In other words, that virtuality helps us understand and interact with others and the real by creating signs – some of the most powerful cognitive artifacts the mind uses to face a reality that would otherwise probably be meaningless.

Thank you for your attention.

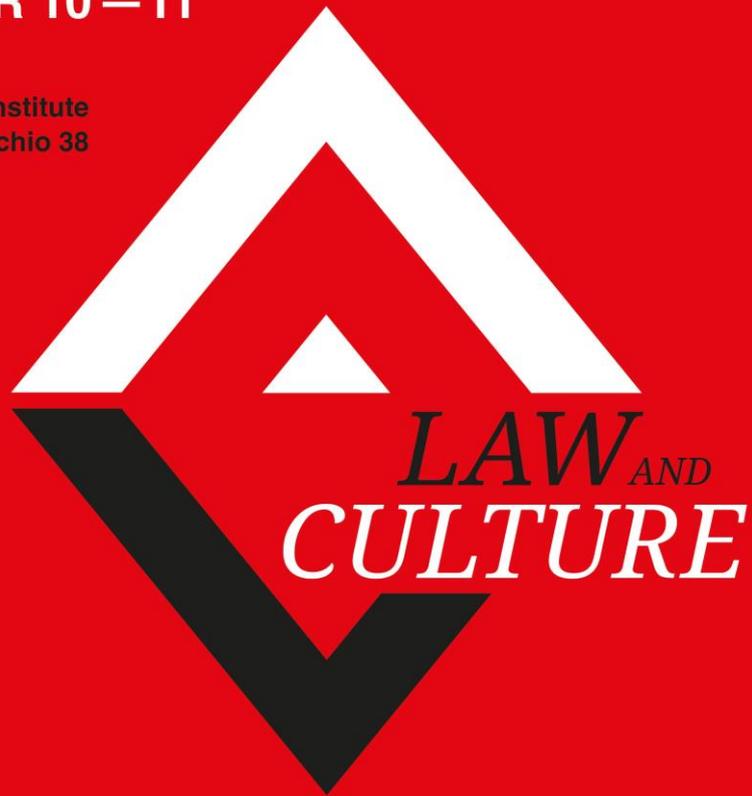